\documentclass{IEEElmagForArxiv}

\usepackage[colorlinks,urlcolor=blue,linkcolor=blue,citecolor=blue]{hyperref}

\usepackage[hyphenbreaks]{breakurl}
\usepackage{braket} 
\usepackage{array}
\usepackage{siunitx}

\usepackage{amssymb,amsmath}

\usepackage{graphicx}

\usepackage{CJKutf8}

\jvol{XX}
\jnum{XX}
\pubyear{2022}

\begin{document}

\sptitle{Microwave Magnetics}

\title{Time-domain two-magnon interference enabled by a tunable beamsplitter}

\author{Cody A. Trevillian\affilmark{1}*} 
\author{Steven Louis\affilmark{2}*}
\author{Vasyl Tyberkevych\affilmark{1}*}

\affil{Department of Physics, Oakland University, Rochester, MI 48309, USA}
\affil{Department of Electrical and Computer Engineering, Oakland University, Rochester, MI 48309, USA}

\IEEEmember{*Member, IEEE}

\corresp{Corresponding authors: Cody A. Trevillian (trevillian@oakland.edu) }

\markboth{Preparation of Papers for \emph{IEEE Magnetics Letters}}{Author Name}

\begin{abstract}
This letter presents a model system for controllable two-magnon interference in the time domain.
This two-magnon interference, i.e., a magnonic analog to the photonic Hong-Ou-Mandel effect, is supported by a tunable magnonic beamsplitter operation formed in a hybrid cavity magnonic system comprising a pair of mutually coupled magnon modes. 
By applying a time-dependent magnetic field, magnons can be excited independently in each mode and subsequently brought into interaction, shifting from independent to collective oscillations, to realize a controllable magnonic beamsplitter.
When the beamsplitter operation is applied to an initially unentangled two-magnon state, a maximally entangled magnonic $N00N$ state with tunable phase sensitivity is produced.
These findings suggest that two-magnon interference in hybrid cavity magnonic systems may enable novel quantum metrological devices to study fundamental magnon dynamics and contribute to developing hybrid magnonic quantum computing architectures.

\end{abstract}


\maketitle

\section{INTRODUCTION}

Since its first demonstration in quantum optics {[}Hong, 1987{]}, Hong–Ou–Mandel (HOM) interference has become a foundational phenomenon in quantum information science {[}Bouchard 2020{]}, enabling critical applications in entanglement generation, quantum computing, and quantum metrology, among others {[}Kapale 2004, Dowling 2008, Dowling 2015{]}.
Beyond optics, two-particle interference has also been observed in systems of bosonic and even fermionic quantum objects, including electrons, atoms, and phonons, reflecting a broader push toward hybrid quantum platforms {[}Bouchard 2020{]}.
Despite this progress, the realization of HOM interference in magnonic systems remains largely underdeveloped, having only recently received attention {[}Kostylev 2023, Qi 2023, Zhu 2023{]}.
This gap is particularly surprising given the potential of magnons to serve as mediators of quantum information in hybrid quantum systems {[}Lachance-Quirion 2019, Li 2020, Awschalom 2021{]}.

Hybrid quantum systems further broaden the application scope of two-particle interference, particularly by enabling interactions between diverse quantum objects {[}Xiang 2013, Clerk 2020, Kurizki 2015, Li 2020, Awschalom 2021{]}.
Magnons—the quanta of spin-wave excitations {[}Gurevich 1996, Kalinikos 1986, Chumak 2015{]} in ordered magnetic materials—possess several properties in this context that make them compelling candidates for quantum information science and quantum hybrid systems {[}Li 2020, Awschalom 2021, Xu 2021{]}.
They can couple strongly to a wide range of both bosonic and fermionic quantum objects {[}Tabuchi 2014, Zhang 2014, Tabuchi 2015, Li 2019, Li 2020, Awschalom 2021, Xu 2021{]}, including phonons {[}Li 2021{]}, microwave and optical photons {[}Tabuchi 2014, Zhang 2014, Li 2019, Shen 2022{]}, superconducting qubits {[}Tabuchi 2015, Tabuchi 2016{]}, and spin defects {[}Candido 2020{]}, enabling coherent information transfer across disparate platforms {[}Li 2020, Awschalom 2021, Xu 2021{]}.
Furthermore, their GHz-to-THz resonance frequencies can be readily tuned using modest external magnetic fields {[}Tabuchi 2014, Zhang 2014, Tabuchi 2015{]}, making them attractive candidates for quantum control {[}Xu 2021, Song 2023, Song 2025{]}.
These attributes position magnons as promising systems for exploring quantum interference phenomena {[}Song 2023, Song 2025{]} and for extending quantum optical techniques into magnonics {[}Li 2020, Awschalom 2021, Tabuchi 2016{]}.
Although this has motivated recent proposals of Floquet-engineered magnonic $N00N$ states {[}Li 2023, Qi 2023, Zhu 2023{]}, as well as magnonic adaptations of standard HOM designs {[}Kostylev 2023{]}, 
two-magnon interference schemes suitable for cavity magnonics applications remain unexplored.

In this context, magnonic systems can be broadly divided into two regimes: one involving propagating spin waves and the other involving localized, discrete magnon modes.
Cavity magnonics belongs to the latter category, focusing on non-propagating magnon modes coherently coupled to confined electromagnetic fields {[}Rameshti 2022{]}.
In such systems, quantum dynamics are most naturally described in the time domain via tunable mode frequencies and controlled interaction windows, rather than through spatial routing of waves.
While spatial beam splitters are fundamental elements of optical interferometry, analogous static beamsplitting elements do not naturally arise for localized magnon modes.
As a result, interference between cavity magnons must be implemented dynamically through time-dependent control.

From this perspective, a temporal beamsplitter can be defined as a unitary operation that coherently mixes two magnon modes during a finite interaction window {[Mendon\c{c}a 2003, Wang 2023 {]}.
By dynamically tuning the frequency separation
between modes, magnon–magnon coupling can be switched on and off in time, realizing the functional equivalent of a beamsplitter without relying on spatial interference.
This time-domain approach provides a natural and compact framework for implementing two-magnon interference in cavity-based magnonic systems.

In this work, we introduce a general and minimal model system for realizing two-magnon interference using dynamically induced magnon–magnon interactions.
By modulating a time-dependent magnetic field, we control the resonance conditions of two localized magnon modes to implement a tunable magnonic temporal beamsplitter operation capable of generating maximally entangled magnonic $N00N$ states from initially separable states.

The significance of our approach is underscored by ongoing advancements in designs of single-magnon sources {[}Chumak 2021{]} and enhanced fabrication methods that facilitate stronger on-chip magnon–photon coupling within hybrid quantum magnonic architectures {[}Chumak 2021, Baity 2021{]}.
These improved fabrication methods, such as those used in YIG-based systems and planar magnetic heterostructures, have made precise temporal control of magnon interactions increasingly feasible {[}Xu 2021, Awschalom 2021, Song 2025, Song 2023, Rao 2025{]}.
These developments suggest that time-domain control protocols, like the one proposed here, could be implemented with existing or near-future technology {[}Li 2022, Song 2025, Song 2023, Rao 2025{]}.

Our minimal and versatile model system provides a practical pathway for integrating magnon-based quantum elements into emerging quantum computing and metrology platforms, consistent with current efforts toward scalable and integrable quantum technologies.
This work establishes a foundation for exploring quantum interference phenomena in magnonic systems and demonstrates the feasibility of time-domain magnonic interference as a controllable quantum resource.
We demonstrate that entangled magnon states can be controllably generated through a tunable magnonic temporal beamsplitter, with phase tunability emerging naturally from the time-domain interaction.
These findings establish two-magnon interference in the time-domain as a scalable mechanism for quantum control in cavity magnonics and hybrid quantum platforms.

\section{MAGNONIC TEMPORAL BEAMSPLITTER MODEL}

\begin{figure}
\includegraphics[width=\linewidth]{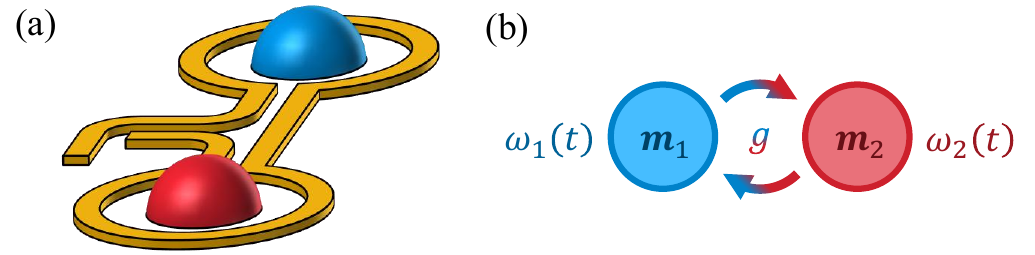}
\caption{\label{fig:candidate_geometries}
Candidate geometries for time-domain magnonic interference.
(a) Representative hybrid magnonic platform in which two spatially separated magnon modes (red, blue) are indirectly coupled via an intermediate bosonic channel (gold).
(b) Generalized model used in this work, consisting of two magnon modes with tunable frequencies $\omega_1(t)$ and $\omega_2(t)$ 
mutually coupled by an effective interaction $g$.
}
\end{figure}

The system considered in this work is motivated by recent experimental proposals and demonstrations of hybrid magnonic platforms, such as those studied by {[}Li 2022, Song 2023, Song 2025{]}, as schematically shown in Fig.~\ref{fig:candidate_geometries}(a). 
In these systems, two spatially distinct magnon modes are indirectly coupled through an intermediate bosonic channel, enabling coherent magnon–magnon interactions. 
Related effective magnon-magnon interactions between localized magnon modes have also been proposed in alternative architectures, including, e.g., synthetic antiferromagnet bilayer systems {[}Wang 2024{]}.

Rather than focusing on a specific physical realization, we adopt a generalized, geometry-agnostic model that captures features common to these platforms, as illustrated in Fig.~\ref{fig:candidate_geometries}(b). 
We consider two localized magnon modes, $m_1$ and $m_2$, with tunable resonance frequencies $\omega_1(t)$ and $\omega_2(t)$ that are mutually coupled with a coherent coupling rate $g$. Such coupling between modes, resulting from an effective interaction of strength $g$, can arise from, e.g., photonic, magnonic, or  other hybrid mediating mechanisms {[}Yi 2022, Wang 2024{]}.
This abstraction enables the study of time-domain magnonic interference independent of any particular geometry or material realization.

The interaction between these two magnon modes can be dynamically controlled through their frequency gap, $\Delta\omega(t) = \omega_1(t) - \omega_2(t)$, using, e.g., tuning by time-dependent external magnetic fields.
When the modes are near resonance ($|\Delta\omega|\lesssim g$), the system is in the strong coupling regime, and coherent magnon-magnon interactions enable quantum information exchange and quantum interference, supporting the formation of quantum entanglement.
Conversely, when the modes are substantially detuned ($|\Delta\omega|\gg g$), the system is in the weak coupling regime, interaction is suppressed and the two modes evolve nearly independently, allowing them to be addressed individually.
This means that by applying shaped magnetic field pulses, one can controllably bring the system into and out of resonance to enable dynamic magnon-magnon interactions.
This is the essential mechanism for achieving time-domain two-magnon interference.
Precisely timed magnetic-field pulses therefore enable controlled switching between interacting and noninteracting regimes, forming the basis of the time-domain two-magnon interference protocol presented here.

To focus on the fundamental mechanisms underlying time-domain two-magnon interference, we adopt an idealized unitary model governed solely by coherent magnon–magnon interactions. 
We neglect non-number-conserving noise processes, such as magnon loss, as well as number-conserving processes, such as phase noise. 
Although relevant in experimental systems, these effects are platform specific and not intrinsic to the geometry-agnostic interference protocol developed here.

To simulate the coherent evolution of our model hybrid magnonic system, the Hamiltonian $\hat{\mathcal{H}}(t)$ for this system is given by
\begin{equation}
\hat{\mathcal{H}}(t)/\hbar = \omega_1(t) \hat{m}_1^\dagger \hat{m}_1 + \omega_2(t) \hat{m}_2^\dagger \hat{m}_2 + g \hat{m}_1^\dagger \hat{m}_2 + g^* \hat{m}_2^\dagger \hat{m}_1,
\label{eq:Hamiltonian}
\end{equation}
where $\hat{m}_{1,2}^\dagger$ ($\hat{m}_{1,2}$) are the creation (annihilation) operators for the magnon modes.
The magnon modes are bosonic quantum objects described by Fock states $\ket{m_{1,2}}$ with a non-negative integer number of magnons $m_{1,2}$.

The phase space of the hybrid magnonic system is spanned by the tensor product states $\ket{m_1, m_2} = \ket{m_1} \otimes \ket{m_2}$.
Note that Eq.~(\ref{eq:Hamiltonian}) conserves the total number of quanta $N = m_1 + m_2$. 
This means that, in absence of any interactions to induce a change of quanta $N$, the dynamics of the system can be considered separately within finite-dimensional subspaces corresponding to different quanta $N$, yielding a block-diagonal decomposition of the Hilbert space.
Here, we restrict our analysis to the subspaces with total number of quanta $N \leq 2$, corresponding to at most one initial excitation in each of the two magnon modes, i.e., $m_{1,2} \leq 1$.
Each excitation manifold corresponds to an invariant subspace with fixed dimension: $ \mathcal{H}_0 $ (1D), $ \mathcal{H}_1 $ (2D), and $ \mathcal{H}_2 $ (3D).
Accordingly, the full Hamiltonian is written as a direct sum (denoted by $\oplus$) over number-conserving subspaces such that
\begin{equation}
    \hat{\mathcal{H}}(t) = \bigoplus_{N=0}^{2}\hat{\mathcal{H}}_{N}(t) = \hat{\mathcal{H}}_{0}(t) \oplus \hat{\mathcal{H}}_{1}(t) \oplus \hat{\mathcal{H}}_{2}(t).
    \label{eq:subspaces}
\end{equation}
This means that the state of the system $\ket{\psi(t)}$ is given by a sum over the states of each $N$-subspace $\ket{\psi_{N}(t)}$
\begin{equation}
    \ket{\psi(t)} = \sum_{N=0}^{2}\ket{\psi_{N}(t)} = \ket{\psi_{0}(t)} + \ket{\psi_{1}(t)} + \ket{\psi_{2}(t)},
    \label{eq:N_state}
\end{equation}
where each $\ket{\psi_{N}(t)}$ can be written as
\begin{equation}
    \ket{\psi_{N}(t)} = \sum_{j + k = N} c_{j,k}(t) \ket{j,k}.
\end{equation}

The state of the system evolves according to the time-dependent Schr\"odinger equation
\begin{equation}
\frac{\partial}{\partial t}\ket{\psi(t)} =
-\frac{i}{\hbar}\hat{\mathcal{H}}(t)\ket{\psi(t)}.
\label{eq:schrodinger}
\end{equation}
The formal solution of Eq.~(\ref{eq:schrodinger}) can be written in terms of the unitary time-evolution operator $\hat{\mathcal{U}}(t,t_0)$, which describes the evolution of $\ket{\psi(t)}$ under influence of $\hat{\mathcal{H}}(t)$ from time $t_0$ to time $t$:
\begin{equation}
\ket{\psi(t)} = \hat{\mathcal{U}}(t,t_0)\ket{\psi(t_0)}
= \mathcal{T}\exp\!\left[-\frac{i}{\hbar}
\int_{t_0}^{t}\hat{\mathcal{H}}(s)\,ds\right]\ket{\psi(t_0)},
\label{eq:unitary}
\end{equation}
where $\mathcal{T}$ denotes the time-ordering operation.

In general, $\hat{\mathcal{U}}(t,t_0)$ may be obtained numerically, for example by using the Magnus expansion {[}Blanes 2009{]}.
In the present implementation, however, we consider the case where the magnetic-field pulse is chosen to be rectangular, with a duration $\tau$, so that $\Delta\omega$ remains constant during each interaction window. 
The Hamiltonian is therefore time-independent over that interval, allowing the time-evolution operator to be written in matrix exponential form as
\begin{equation}
\hat{\mathcal{U}}(t,t_0)
= \exp\left[-\frac{i}{\hbar}\hat{\mathcal{H}}[\Delta\omega]
\,\tau
\right]
\equiv \hat{\mathcal{U}}_0(\Delta\omega,\tau).
\label{eq:U0}
\end{equation}
The resulting evolution therefore depends only on the frequency gap $\Delta\omega$ and the interaction time $\tau=t-t_0$.

\section{Temporal Beamsplitter Protocol and Demonstration}

In conventional interferometry, a beam splitter is a static optical element that coherently mixes two spatial modes, producing a superposition of transmitted and reflected paths.
Such devices rely on fixed coupling geometry in space to implement a unitary transformation acting on mode states.
A temporal beam splitter performs the same fundamental operation, but replaces spatial mixing with a dynamically controlled interaction that is switched on and off in time.
From this perspective, both spatial and temporal beam splitters are best understood as unitary operations acting on mode states, with the distinction arising from whether the coupling is realized through static geometry or through time-dependent control of the system Hamiltonian.

\begin{figure}
\includegraphics[width=\linewidth]{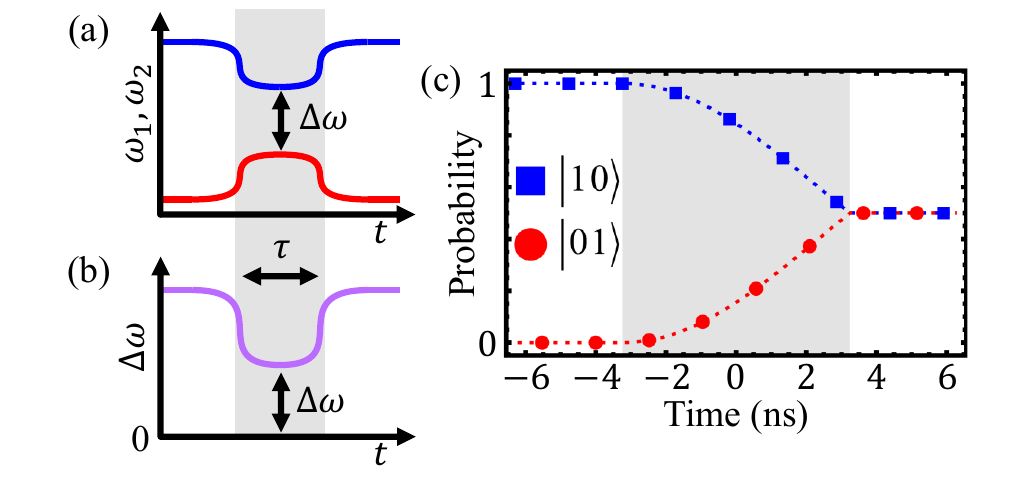}
\caption{\label{fig:temporal_BS}
Single-magnon characterization of the temporal magnonic beamsplitter.
(a) Time-dependent magnon frequencies $\omega_1(t)$ (blue) and $\omega_2(t)$ (red) are tuned by a magnetic-field pulse, reducing $\Delta\omega$ for a duration $\tau$ (gray).
(b) Temporal profile of $\Delta\omega$.
(c) Simulated evolution of an $\ket{10}$ state under a balanced TBS pulse to $(\ket{10}+\ket{01})/\sqrt{2}$.
Simulation parameters: $\Delta\omega=0$, $g=2\pi\times20$ MHz.
}
\end{figure}

In the present framework, the temporal beamsplitter (TBS), denoted $\hat{\mathcal{U}}_{\mathrm{TBS}}^{(\Delta\omega)}$, corresponds to a specific realization of (\ref{eq:U0}).
The temporal beamsplitter is designed such that an initial single-magnon state evolves according to
\begin{equation}
\hat{\mathcal{U}}_{\mathrm{TBS}}^{(\Delta\omega)}\ket{10}
= \frac{1}{\sqrt{2}}\bigg(\ket{10}+e^{i\varphi}\ket{01}\bigg).\label{asdefw}
\end{equation} 
Here, $\varphi$ is the relative phase accumulated between the two magnon modes during the time evolution.
This operation is implemented by selecting $\Delta\omega$ and $\tau$, as shown in Figs.~\ref{fig:temporal_BS}(a),(b), such that the modes are brought into resonance ($|\Delta\omega|\lesssim g$) for a finite interaction window of duration $\tau$ during which coherent exchange occurs.

The parameters $\Delta\omega$ and $\tau$ are selected such that the transition probability of a single magnon between modes is $50\%$. 
That is, if the system is initialized in the single-magnon state $\ket{10}$, the TBS pulse evolves it into the coherent superposition $\alpha\ket{10}+\beta\ket{01}$.
Selecting $\Delta\omega$ and $\tau$ such that $|\alpha|=|\beta|=1/\sqrt{2}$ defines a balanced magnonic temporal beamsplitter.

The coherent exchange of a single magnon between the two modes proceeds via Rabi oscillations with angular frequency $(1/2)\sqrt{(2g)^2+\Delta\omega^2}$.
A balanced beamsplitter, with a tunneling probability of $1/2$, is obtained by interrupting these oscillatory dynamics halfway through a full population transfer between the modes.

In the resonant limit, $\Delta\omega=0$, the magnon modes undergo the shortest coherent exchange dynamics, and a balanced temporal beamsplitter is obtained for an interaction time $\tau=\pi/4g$.
For a finite frequency gap, $\Delta\omega\neq 0$, the same beamsplitting condition can be achieved by increasing the interaction time, compensating for the reduced effective coupling between the modes.

Figure~\ref{fig:temporal_BS}(c) presents a numerical demonstration of the TBS through the simulated time evolution of a system initialized in the single-magnon state $\ket{10}$, with the two magnon modes ($m_1$ blue, $m_2$ red) initially far detuned and therefore noninteracting.
A pulsed magnetic field is then applied to dynamically reduce the frequency gap, bringing the modes into resonance for a duration $\tau$, as indicated by the shaded gray region.
During this interaction window, coherent magnon exchange occurs due to the direct mode coupling, resulting in the final superposition state according to (\ref{asdefw}). 
This single-magnon result verifies that the temporal beamsplitter implements the intended 50:50 unitary mode-mixing operation.

\section{Two-Magnon Interference and Entanglement Generation}

\begin{figure}
\includegraphics[width=\linewidth]{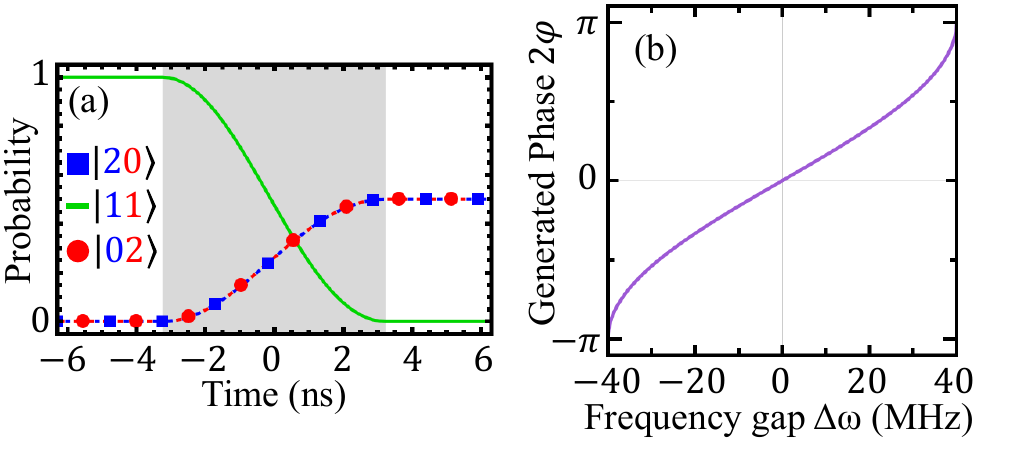}
\caption{\label{fig:generation_detection}
Controllable 
generation 
of maximally entangled 
magnonic $N00N$ states using a TBS pulse.
(a) Simulated time evolution showing coherent transformation from $ |11\rangle $ to the entangled state $ (|20\rangle + e^{i2\varphi}|02\rangle)/\sqrt{2} $.
(b) Output phase $ \varphi $ of the $N00N$ state as a function of $ \Delta\omega $.
}
\end{figure}

Having characterized the action of the TBS in the single-magnon limit, we now consider its behavior when two magnons are present in the system in an initially unentangled state.
This regime gives rise to qualitatively new physics, as the same unitary operation produces genuine two-particle interference between otherwise non-interacting and localized bosons.

In the simulation, the system is initialized in the separable state $\ket{11}$, corresponding to one magnon in each mode, as shown in Fig.~\ref{fig:generation_detection}(a).
This configuration is the direct analog of injecting one particle into each input port of an optical beamsplitter.
The same TBS pulse characterized in the single-magnon simulations is then applied during the interaction window (gray region), with identical control parameters.
When the $\tau$ is chosen to realize a balanced beamsplitter, $\hat{\mathcal{U}}_\mathrm{TBS}^{(\Delta\omega)}$ transforms the initial state according to
\begin{equation}
\hat{\mathcal{U}}_\mathrm{TBS}^{(\Delta\omega)} \ket{11}
=
\frac{1}{\sqrt{2}}
\left(
\ket{20} + e^{i2\varphi} \ket{02}
\right).
\end{equation}
The output is therefore a maximally-path-entangled two-magnon $N00N$ state.

This behavior, accompanied by the characteristic dip of the $\ket{11}$ probability, reflects a two-particle bosonic interference of magnons. 
Under balanced beamsplitting, destructive interference suppresses the $\ket{11}$ state, converting it into a coherent superposition of $\ket{20}$ and $\ket{02}$, in direct anology with HOM interference. 
Here the TBS pulse replaces the optical beamsplitter, providing mode mixing in time rather than space.

This result captures the central physical insight of the present work.
An initially unentangled two-magnon state is controllably converted into an entangled state through a single time-domain control operation.
No ancillary systems, measurements, or nonlinear interactions are required.
Temporal control alone is sufficient to generate magnon–magnon entanglement between otherwise independent localized modes, such as those encountered in cavity magnonic systems.

Beyond state generation, the temporal beamsplitter also imprints a well-defined relative phase between the two components of the $N00N$ state.
The phase $2\varphi$ appearing in the superposition is coherently accumulated during the interaction window.
Figure~\ref{fig:generation_detection}(b) shows the resulting phase accumulated as a function of $\Delta\omega$ applied during the beamsplitter pulse.
As demonstrated in the figure, the relative phase of the two-magnon $N00N$ state is directly controlled by $\Delta \omega$ during the interaction.

\section{CONCLUSION}

We have proposed and analyzed a general, geometry-agnostic protocol for time-domain two-magnon interference based on a tunable magnonic temporal beamsplitter.
By dynamically controlling the frequency gap between two coherently coupled, localized magnon modes, the protocol enables controllable generation of maximally-entangled magnonic $N00N$ states using only direct magnon–magnon interactions applied dynamically in time.
This establishes a magnonic analog of the Hong–Ou–Mandel interference realized entirely in the time domain, without the need for ancillary qubits or complex spatial interferometric structures.

Our results show that in systems composed of non-propagating cavity modes, interference can be realized through programmable temporal control rather than static spatial mixing.
Numerical simulations confirm two-magnon interference and entanglement generation within a geometry-independent model applicable to coupled magnonic systems, including cavity magnonic and cavity 
electromagnonic architectures.

As experimental capabilities in single-magnon generation, strong coupling mechanisms, and on-chip cavity integration continue to advance, temporal beamsplitting protocols such as the one proposed here offer a scalable pathway toward interference-based quantum state control in cavity magnonics and hybrid cavity quantum systems.



\begin{thebibliography}{}

\bibitem{bibAwschalom2021}
Awschalom D. D., Du C. R., He R., Heremans F. J., Hoffmann A., Hou J., Kurebayashi H., Li Y., Liu L., Novosad V. \emph{et al.} (2021),
``Quantum engineering with hybrid magnonic systems and materials,''
\emph{IEEE Trans. Quantum Eng.,} vol. 2, pp. 1--36.

\bibitem{bibBaity2021}
Baity P. G., Bozhko D. A., Mac\^edo R., Smith W., Holland R. C., Danilin S., Seferai V., Barbosa J., Peroor R. R., Goldman S. \emph{et al.} (2021),
``Strong magnon--photon coupling with chip-integrated YIG in the zero-temperature limit,'' 
\emph{Appl. Phys. Lett.,} vol. 119, no. 3.

\bibitem{bibBlanes2009}
Blanes S., Casas F., Oteo J.A., Ros J. (2009),
``The Magnus expansion and some of its applications,'' \emph{Phys. Rep.}, vol. 470, iss. 5–-6, pp 151--238.

\bibitem{bibBouchard2020} 
Bouchard F., Sit A., Zhang Y., Fickler R., Miatto F. M., Yao Y., Sciarrino F., Karimi E. (2020), 
``Two-photon interference: the Hong--Ou--Mandel effect,''
\emph{Rep. Prog. Phys.,} vol. 84, no. 1, p. 012402.

\bibitem{bibCandido2020}
Candido D. R., Fuchs G. D., Johnston-Halperin E., Flatt\'e M. E. (2020),
``Predicted strong coupling of solid-state spins via a single magnon mode,''
\emph{Mater. Quantum Technol.,} vol. 1, no. 1, p. 011001.

\bibitem{bibClerk2020}
Clerk A. A., Lehnert K. W., Bertet P., Petta J. R., Nakamura Y. (2020),
``Hybrid quantum systems with circuit quantum electrodynamics,''
\emph{Nat. Phys.,} vol. 16, no. 3, pp. 257--267.

\bibitem{bibChumak2015}
Chumak A. V., Vasyuchka V. I., Serga A. A., Hillebrands B. (2015),
``Magnon spintronics,''
\emph{Nat. Phys.,} vol. 11, no. 6, pp. 453--461.

\bibitem{bibChumak2022}
Chumak A. V., Kabos P., Wu M., Abert C., Adelmann C., Adeyeye A. O., \AA kerman J., Aliev F. G., Anane A., Awad A. \emph{et al.} (2022),
``Advances in magnetics roadmap on spin-wave computing,''
\emph{IEEE Trans. Magn.,} vol. 58, no. 6, pp. 1--72.

\bibitem{bibDowling2008}
Dowling J. P. (2008), ``Quantum optical metrology -- the lowdown on
high-$N00N$ states,'' \emph{Contemp. Phys.,} vol. 49, no. 2,
pp. 125--143.

\bibitem{bibDowlingSeshadreesan2015}
Dowling J. P., Seshadreesan K. P. (2015), 
``Quantum optical technologies for metrology, sensing, and imaging,'' 
\emph{J. Lightw. Technol.,} vol. 33, no. 12, pp. 2359--2370.

\bibitem{bibGurevich1996}
Gurevich A. G., Melkov G. A. (1996),
\emph{Magnetization Oscillations and Waves.}
London, U.K.: CRC Press.

\bibitem{bibHong1987}
Hong C. K. (1987), 
``Measurement of subpicosecond time intervals between two photons by interference,'' 
\emph{Phys. Rev. Lett.,} vol. 59, no. 18, pp. 2044--2046.

\bibitem{bibKalinikos1986}
Kalinikos B. A., Slavin A. N. (1986),
``Theory of dipole-exchange spin wave spectrum for ferromagnetic films with mixed exchange boundary conditions,''
\emph{J. Phys. C: Solid State Phys.,} vol. 19, no. 35, pp. 7013--7033.

\bibitem{bibKapale2004}
Kapale K. T., DiDomenico L. D., Lee H., Kok P., Dowling J. P. (2004),
``Quantum interferometric sensors,'' in \emph{Quantum Sensing and
Nanophotonic Devices,} vol. 5359, SPIE, pp. 169--176.

\bibitem{bibKostylev2023}
Kostylev M. (2023), 
``Magnonic Hong--Ou--Mandel effect,''
\emph{Phys. Rev. B,} vol. 108, no. 13, p. 134416.

\bibitem{bibKurizki2015}
Kurizki G., Bertet P., Kubo Y., M\o lmer K., Petrosyan D., Rabl P., Schmiedmayer J. (2015),
``Quantum technologies with hybrid systems,''
\emph{Proc. Natl. Acad. Sci. U.S.A.,} vol. 112, no. 13, pp. 3866--3873.

\bibitem{bibLachanceQuirion2019}
Lachance-Quirion D., Tabuchi Y., Gloppe A., Usami K., Nakamura Y. (2019),
``Hybrid quantum systems based on magnonics,''
\emph{Appl. Phys. Express,} vol. 12, no. 7, p. 070101.

\bibitem{bibLi2019}
Li Y., Polakovic T., Wang Y.-L., Xu J., Lendinez S., Zhang Z., Ding J., Khaire T., Saglam H., Divan R. \emph{et al.} (2019),
``Strong coupling between magnons and microwave photons in on-chip ferromagnet--superconductor thin-film devices,''
\emph{Phys. Rev. Lett.,} vol. 123, no. 10, p. 107701.

\bibitem{bibLi2020}
Li Y., Zhang W., Tyberkevych V., Kwok W.-K., Hoffmann A., Novosad V. (2020),
``Hybrid magnonics: Physics, circuits, and applications for coherent information processing,'' 
\emph{J. Appl. Phys.,} vol. 128, no. 13, p. 130902.

\bibitem{bibLi2021APLMat}
Li Y., Zhao C., Zhang W., Hoffmann A., Novosad V. (2021),
``Advances in coherent coupling between magnons and acoustic phonons,''
\emph{APL Mater.,} vol. 9, no. 6.

\bibitem{bibLi2022}
Li Y., Yefremenko V. G., Lisovenko M., Trevillian C., Polakovic T., Cecil T. W., Barry P. S., Pearson J., Divan R., Tyberkevych V. \emph{et al.} (2022),
``Coherent coupling of two remote magnonic resonators mediated by superconducting circuits,'' 
\emph{Phys. Rev. Lett.,} vol. 128, no. 4, p. 047701.

\bibitem{bibMendonca2003}
Mendon\c{c}a J. T., Martins A. M., Guerreiro A. (2003),
``Temporal beam splitter and temporal interference,''
\emph{Phys. Rev. A,} vol. 68, no. 4, p. 043801.

\bibitem{bibQi2023}
Qi S., Jing J. (2023), 
``Floquet generation of a magnonic NOON state,''
\emph{Phys. Rev. A,} vol. 107, no. 1.

\bibitem{bibRameshti2022}
Rameshti B. Z., Kusminskiy S. V., Haigh J. A., Usami K., Lachance-Quirion D., Nakamura Y., Hu C.-M., Tang H. X., Bauer G. E. W., Blanter Y. M. (2022),
``Cavity magnonics,'' 
\emph{Phys. Rep.,} vol. 979, pp. 1--61.

\bibitem{bibRao2025}
Rao J., Wang Y.-P., Chen Z., Yao B., Zhao K., Wei C., Wang C., Li R., Bai L., Lu W. (2025),
``Time-varying strong coupling and the induced time diffraction of magnon modes,'' 
\emph{Phys. Rev. Lett.,} vol. 135, no. 6, p. 066704.

\bibitem{bibShen2022}
Shen Z., Xu G.-T., Zhang M., Zhang Y.-L., Wang Y., Chai C.-Z.,
Zou C.-L., Guo G.-C., Dong C.-H. (2022),
``Coherent coupling between phonons, magnons, and photons,''
\emph{Phys. Rev. Lett.,} vol. 129, no. 24, p. 243601.

\bibitem{bibSong2025}
Song M., Polakovic T., Lim J., Cecil T. W., Pearson J., Divan R., Kwok W.K., Welp U., Hoffmann A., Kim K.J. \emph{et al.} (2025),
``Single-shot magnon interference in a magnon-superconducting-resonator hybrid circuit,'' 
\emph{Nat. Commun.,} vol. 16, no. 1, p. 3649.

\bibitem{bibSong2023arXiv}
Song M., Polakovic T., Lim J., Cecil T. W., Pearson J., Divan R., Kwok W.-K., Welp U., Hoffmann A., Kim K.J. \emph{et al.} (2023),
``Programmable real-time magnon interference in two remotely coupled magnonic resonators,'' 
arXiv:2309.04289.

\bibitem{bibTabuchi2016}
Tabuchi Y., Ishino S., Noguchi A., Ishikawa T., Yamazaki R., Usami K.,
Nakamura Y. (2016),
``Quantum magnonics: The magnon meets the superconducting qubit,''
\emph{C. R. Phys.,} vol. 17, no. 7, pp. 729--739.

\bibitem{bibTabuchi2014}
Tabuchi Y., Ishino S., Ishikawa T., Yamazaki R., Usami K., Nakamura Y. (2014),
``Hybridizing ferromagnetic magnons and microwave photons in the quantum limit,'' 
\emph{Phys. Rev. Lett.,} vol. 113, no. 8, p. 083603.

\bibitem{bibTabuchi2015Science}
Tabuchi Y., Ishino S., Noguchi A., Ishikawa T., Yamazaki R., Usami K., Nakamura Y. (2015),
``Coherent coupling between a ferromagnetic magnon and a superconducting qubit,'' 
\emph{Science,} vol. 349, no. 6246, pp. 405--408.

\bibitem{bibWang2023}
Wang S., Qin C., Zhao L., Ye H., Longhi S., Lu P., Wang B. (2023),
``Photonic Floquet Landau-Zener tunneling and temporal beam splitters,''
\emph{Sci. Adv.,} vol. 9, iss. 18, no. eadh0415.

\bibitem{bibWang2024}
Wang Y., Zhang Y.
Li C., Wei J., He B., Xu H., Xia J., Luo X., Li J., Dong J. \emph{et al.} (2024), ``Ultrastrong to nearly deep-strong magnon-magnon coupling with a high degree of freedom in synthetic antiferromagnets,'', \emph{Nat. Comm.}, vol. 15, iss. 1, no. 2077.

\bibitem{bibXiang2013}
Xiang Z.L., Ashhab S., You J. Q., Nori F. (2013), 
``Hybrid quantum circuits: Superconducting circuits interacting with other quantum systems,'' 
\emph{Rev. Mod. Phys.,} vol. 85, no. 2, pp. 623--653.

\bibitem{bibXu2021}
Xu J., Zhong C., Han X., Jin D., Jiang L., Zhang X. (2021),
``Coherent gate operations in hybrid magnonics,''
\emph{Phys. Rev. Lett.,} vol. 126, no. 20, p. 207202.

\bibitem{bibZhang2014}
Zhang X., Zou C.-L., Jiang L., Tang H. X. (2014),
``Strongly coupled magnons and cavity microwave photons,''
\emph{Phys. Rev. Lett.,} vol. 113, no. 15, p. 156401.

\bibitem{bibZhu2023}
Zhu X., Xia R., Xu L. (2023), 
``Floquet-engineering magnonic NOON states with performance improved by soft quantum control,''
\emph{Quantum Inf. Process.,} vol. 22, no. 12, pp. 1--15.

\end{thebibliography}
\end{document}